# Regge behaviour of distribution functions and t and x-evolutions of gluon distribution function at low-x

U. Jamil[1] and J.K. Sarma[2]

Department of Physics, Tezpur University, Napaam, Tezpur-784028, Assam, India

**Abstract.** In this paper t and x-evolutions of gluon distribution function from Dokshitzer-Gribov-Lipatov-Altarelli-Parisi (DGLAP) evolution equation in leading order (LO) at low-x, assuming the Regge behaviour of quark and gluon at this limit, are presented. We compare our results of gluon distribution function with MRST 2001, MRST 2004 and GRV '98 parameterizations and show the compatibility of Regge behaviour of quark and gluon distribution functions with perturbative quantum chromodynamics (PQCD) at low-x. We also discuss the limitations of Taylor series expansion method used earlier to solve DGLAP evolution equations, in the Regge behaviour of distribution functions.

## 1 Introduction

The measurement of proton and deuteron structure functions by deep inelastic scattering (DIS) processes in the low-x region, where x is the Bjorken variable, have been of great importance for the understanding of quark-gluon substructure of hadrons [1, 2]. In addition to this, it is also important to have the knowledge of the gluon distribution inside hadrons at low-x, because gluons are expected to be dominant in that region. Moreover, gluon distributions are important inputs in many high energy processes and also important for examination of PQCD [3], the underlying dynamics of quarks and gluons. In PQCD, the high-$Q^2$ behaviour of DIS can be explained by the DGLAP evolution equations [4, 5]. These equations introduced the parton distribution functions which can be interpreted as the probability of finding, say in a proton, respectively a quark, an antiquark or a gluon with virtuality less than $Q^2$ and with momentum fraction x. When $Q^2 \to \infty$ i.e., very large, the $Q^2$-evolution of these densities (at fixed-x) are given by the DGLAP equations and these are often considered as a very good test of PQCD. The solutions of the DGLAP equations can be calculated either by numerical integration in steps or by taking the moments of the distributions [6]. The approximate solutions of DGLAP evolution equations have been reported in recent years [7−11] with considerable phenomenological success. And structure functions thus calculated are expected to rise approximately with a power of x towards low-x which is supported by Regge theory [12, 13].

---

[1]jamil@tezu.ernet.in, [2]jks@tezu.ernet.in

The low-x region of DIS offers a unique possibility to explore the Regge limit [12] of PQCD. The low-x behaviour of parton distributions can be considered by a triple pole pomeron model [13, 14] at the initial scale $Q_0^2$ and then evolved using DGLAP equations. It is possible to have the same singularity structure for quarks and gluons, in agreement with Regge theory. The Regge behaviour of the structure function $F_2$ in the high-$Q^2$ region reflects itself in the low-x behaviour of the quark and antiquark distributions. Thus the Regge behaviour of the sea quark and antiquark distributions for low-x is given by $q_{sea}(x) \sim x^{\lambda_p}$ with pomeron exchange [13] of intercept $\lambda_p = -1$. But the valence quark distribution for low-x given by $q_{val}(x) \sim x^{-\lambda_r}$ corresponding to a reggeon exchange of intercept $\lambda_r = 1/2$.

In our present work, we have derived the solution of DGLAP evolution equation for gluon distribution function at low-x in leading order (LO) considering Regge behaviour of distribution functions. The t and x-evolutions of LO gluon distribution functions thus obtained have been compared with global MRST and GRV '98 parameterizations. Here we overcome the limitations that arise from Taylor series expansion method [15, 16] and this method is also mathematically simple. In this paper, section 1, section 2, section 3 and section 4 are the introduction, theory, results and discussions and conclusions respectively.

## 2 Theory

The LO DGLAP evolution equation for gluon distribution function has the standard form [17, 18]

$$\frac{\partial G(x, t)}{\partial t} - \frac{A_f}{t}\left\{\left(\frac{11}{12} - \frac{N_f}{18} + \ln(1-x)\right)G(x, t) + I_g\right\} = 0, \tag{1}$$

where

$$I_g = \int_x^1 d\omega \left[\frac{\omega G(x/\omega, t) - G(x, t)}{1-\omega} + \left(\omega(1-\omega) + \frac{1-\omega}{\omega}\right)G(x/\omega, t) + \frac{2}{9}\left(\frac{1+(1-\omega)^2}{\omega}\right)F_2^s(x/\omega, t)\right],$$

$t = \ln(Q^2/\Lambda^2)$, $\Lambda$ being the QCD cut-off parameter and $A_f = 4/(33-2N_f)$, $N_f$ being the number of flavours.

As the gluons are expected to be dominant at low-x, so we can neglect the quark contribution to the evolution equation of gluon distribution function and we can phenomenologically get the amount of contribution of quark to the gluon distribution function at different x and $Q^2$.

Among the various methods to solve DGLAP equations, one simple method is to use Taylor expansion [19] to transform the integro-differential equations into partial differential equations and thus to solve them by standard methods [20, 21]. But when we consider Regge behaviour of structure functions to solve these evolution equations, the use of Taylor expansion becomes limited. In this method, we introduce the variable $u = 1 - \omega$ and we get $\dfrac{x}{\omega} = \dfrac{x}{1-u} = x\sum_{k=0}^{\infty} u^k$.

Since $0 < u < (1-x)$, $|u| < 1$, it implies that $\dfrac{x}{1-u} = x\sum_{k=0}^{\infty} u^k$ is convergent. Applying the Taylor expansion, for example, for the gluon distribution function, we get

$$G\left(\frac{x}{\omega}, t\right) = G\left(\frac{x}{1-u}, t\right) = G\left(x + x\sum_{k=1}^{\infty} u^k, t\right)$$

$$= G(x, t) + x\sum_{k=1}^{\infty} u^k \frac{\partial G(x, t)}{\partial x} + \frac{1}{2} x^2 \left(\sum_{k=1}^{\infty} u^k\right)^2 \frac{\partial^2 G(x, t)}{\partial x^2} + \ldots\ldots \quad (2)$$

When we apply the Regge behaviour, we take the form of gluon distribution function as $G(x, t) = A(t) x^{-\lambda}$, where $A(t)$ is a function of $t$, and $\lambda$ is the intercept. Then

$$\frac{\partial G(x,t)}{\partial x} = A(t)(-\lambda)x^{-\lambda-1} = A(t)x^{-\lambda}(-\lambda)x^{-1} = (-1)\lambda x^{-1} G(x,t),$$

$$\frac{\partial^2 G(x,t)}{\partial x^2} = A(t)(-\lambda)(-\lambda-1)x^{-\lambda-2} = A(t)x^{-\lambda}(-\lambda)x^{-1} = (-1)^2 \lambda(\lambda+1)x^{-2} G(x,t),$$

$$\frac{\partial^3 G(x,t)}{\partial x^3} = (-1)^3 \lambda(\lambda+1)(\lambda+2)x^{-3} G(x,t) \text{ and so on. So equation (2) becomes}$$

$$G\left(\frac{x}{\omega}, t\right) = G(x,t) + \left(\sum_{k=1}^{\infty} u^k\right)(-1)\lambda G(x,t) + \frac{1}{2}\left(\sum_{k=1}^{\infty} u^k\right)^2 (-1)^2 \lambda(\lambda+1) G(x,t) + \ldots\ldots \quad .$$

In the expansion series, we will get terms with alternate positive and negative signs and contribution from $\lambda$ to each term increases. So in this case, it is not possible to truncate this infinite series in to finite number of terms by applying boundary condition such as low-x [22] or so and also this is not a convergent series [19]. So, in solving DGLAP evolution equation applying Regge behaviour of distribution functions, we cannot apply Taylor series expansion method.

Now let us consider the Regge behaviour of singlet structure function [13, 23] as

$$F_2^s(x, t) = T(t) x^{-\lambda_s}, \quad (3)$$

where $T(t)$ is a function of $t$, and $\lambda_s$ is the Regge intercept for singlet structure function. But according to Regge theory, the high energy (low-x) behaviour of both gluons and sea quarks is controlled by the same singularity factor in the complex angular momentum plane [13], and so we would expect $\lambda_s = \lambda_g = \lambda$, where $\lambda$ is taken as a constant factor throughout the calculation. So the equation (3) becomes

$$F_2^s(x, t) = T(t) x^{-\lambda}. \quad (4)$$

From equation (4) we get,

$$F_2{}^s(x/\omega, t) = T(t)\omega^\lambda x^{-\lambda}. \tag{5}$$

Let us assume for simplicity [24, 25],

$$G(x, t) = K(x) F_2{}^s(x, t), \tag{6}$$

where K(x) is a parameter to be determined from experimental data and we assume $K(x) = k$, $ax^b$ or $ce^{dx}$ where k, a, b, c and d are constants. Therefore,

$$G(x/\omega, t) = K(x/\omega) F_2{}^s(x/\omega, t) = K(x/\omega) T(t)\omega^\lambda x^{-\lambda}. \tag{7}$$

Putting equations (5) to (7) in equation (1), we get,

$$\frac{\partial T(t)}{\partial t} - \frac{T(t)}{t} P(x) = 0, \tag{8}$$

where $P(x) = A_f \left\{ A + \ln(1-x) + \frac{f(x)}{K(x)} \right\}$, $A = \frac{11}{12} - \frac{N_f}{18}$ and

$$f(x) = \int_x^1 d\omega \left[ \frac{\left(K(x/\omega)\omega^{\lambda+1} - K(x)\right)}{1-\omega} + \left(\omega(1-\omega) + \frac{1-\omega}{\omega}\right) K(x/\omega)\omega^\lambda + \frac{2}{9}\left(\frac{1 + (1-\omega)^2}{\omega}\right)\omega^\lambda \right].$$

We can solve equation (8) by the standard method of separation of variables [26]. Now equation (8) gives

$$\frac{\partial T(t)}{\partial t} = \frac{T(t)}{t} P(x) \Rightarrow \frac{\partial T(t)}{T(t)} \frac{t}{\partial t} = P(x).$$

As each side of the above equation is a function of different independent variables, so each side must be equal to a constant, V (say).

Therefore, $\dfrac{\partial T(t)}{T(t)} \dfrac{t}{\partial t} = V$ \hfill (9)

and $\quad P(x) = V.$ \hfill (10)

Integrating equation (9) we get,

$\ln T(t) = V \ln t + \ln C$

$\Rightarrow \ln T(t) = \ln(Ct^V)$

$\Rightarrow T(t) = Ct^V.$ \hfill (11)

Therefore, equation (6) becomes,

$$G(x, t) = K(x) C t^V x^{-\lambda}. \tag{12}$$

From equation (10) we can put the value of the constant $V = P(x)$, and equation (12) becomes,

$$G(x, t) = K(x) C t^{P(x)} x^{-\lambda}. \tag{13}$$

At some lower value of $t = t_0$, equation (13) becomes,

$$G(x, t_0) = K(x) C t_0{}^{P(x)} x^{-\lambda}. \tag{14}$$

From equations (13) and (14) we get,

$$G(x, t) = G(x, t_0)(t/t_0)^{P(x)}, \qquad (15)$$

which gives the t-evolution of gluon distribution function in LO.

Now, at $x = x_0$, from equation (13) we get,

$$G(x_0, t) = K(x_0)Ct^{P(x_0)}x_0^{-\lambda}, \qquad (16)$$

From equations (13) and (16), we get,

$$G(x, t) = G(x_0, t)\frac{K(x)}{K(x_0)}t^{\{P(x) - P(x_0)\}}\left(\frac{x}{x_0}\right)^{-\lambda}, \qquad (17)$$

which gives the x-evolution of gluon distribution function in LO.

Now ignoring the quark contribution to the gluon distribution function we get from the standard DGLAP evolution equation (1)

$$\frac{\partial G(x,t)}{\partial t} - \frac{A_f}{t}\left\{\left(\frac{11}{12} - \frac{N_f}{18} + \ln(1-x)\right)G(x,t) + I'_g\right\} = 0, \qquad (18)$$

where

$$I'_g = \int_x^1 d\omega\left[\frac{\omega G(x/\omega, t) - G(x, t)}{1-\omega} + \left(\omega(1-\omega) + \frac{1-\omega}{\omega}\right)G(x/\omega, t)\right].$$

Now pursuing the same procedure as above, we get,

$$G(x, t) = G(x, t_0)(t/t_0)^{B(x)}, \qquad (19)$$

which gives the t-evolution of gluon distribution function ignoring the quark contribution in LO. Here

$$B(x) = A_f\left\{(A + \ln(1-x)) + \int_x^1 d\omega\left[\frac{(\omega^{\lambda+1} - 1)}{1-\omega} + \left(\omega(1-\omega) + \frac{1-\omega}{\omega}\right)\omega^\lambda\right]\right\}.$$

Again,

$$G(x, t) = G(x_0, t)t^{\{B(x) - B(x_0)\}}\left(\frac{x}{x_0}\right)^{-\lambda}, \qquad (20)$$

which gives the x-evolution of gluon distribution function ignoring the quark contribution in LO.

**3 Results and discussions**

In this paper, we have obtained a new description of t and x-evolutions of gluon distribution function considering Regge behaviour of distribution functions given by equations (15) and (17) respectively. We are also interested to see the contribution of quark to gluon distribution function at low-x and high $Q^2$, theoretically which should decrease for x→0, $Q^2$→ ∞ [2, 27]. We have obtained this description of gluon distribution function ignoring the quark contribution in the evolution equation given by equations (19) and (20). We compare our results of t and x-evolutions of gluon distribution function in LO given by equations (15), (17), (19) and (20) with MRST and GRV '98 parameterizations. We have taken the MRST 2001 fit [28] to the CDFIB data [29] for $Q^2 = 20$ GeV$^2$, in which they obtained the optimum global NLO fit with the starting parameterizations of the partons at $Q_0^2 = 1$ GeV$^2$ given by

$$xg = 123.5x^{1.16}(1-x)^{4.69}(1-3.57x^{0.5}+3.41x) - 0.038x^{-0.5}(1-x)^{10}.$$

The optimum fit corresponds to $\alpha_s(M_z^2) = 0.119$ i.e. $\Lambda_{\overline{MS}}(N_f = 4) = 323$ MeV. We have also taken the MRST 2004 fit [30] to the ZEUS [31] and H1 [32] data with x<0.01 and 2<$Q^2$<500 GeV$^2$ for $Q^2 = 100$ GeV$^2$, in which they have taken the MRST-like parametric form same as for MRST 2001 fit [28] for the starting distribution at $Q_0^2 = 1$ GeV$^2$ given by

$$xg = A_g x^{-\lambda_g}(1-x)^{3.7}(1+\varepsilon_g\sqrt{x}+\gamma_g x) - A x^{-\delta}(1-x)^{10},$$

where the powers of the (1–x) factors are taken from MRST 2001 fit. The $\lambda_g$, $\varepsilon_g$, A and $\delta$ are taken as free parameters. The value of $\alpha_s(M_z^2)$ is taken to be the same as in the MRST 2001 fit. We have taken the GRV '98 parameterization [33] for $10^{-2} \leq x \leq 10^{-5}$ and $20 \leq Q^2 \leq 80$ GeV$^2$, where they used H1 [34] and ZEUS [35] high precision data on $G(x, Q^2)$. They have chosen $\alpha_s(M_z^2) = 0.114$ i.e. $\Lambda_{\overline{MS}}(N_f = 4) = 246$ MeV. The input densities have been fixed using the data sets of HERA [34], SLAC [36], BCDMS [37], NMC [38] and E665 [39]. The resulting input distribution at $Q^2 = 0.40$ GeV$^2$ is given by

$$xg = 20.80x^{1.6}(1-x)^{4.1}.$$

We compare our results from the equations (15) and (17) for $K(x) = k$, $ax^b$ and $ce^{dx}$, where k, a, b, c and d are constants. In our work, we have found the values of the gluon distribution function remain almost same for b<0.0001 and for d>–0.001. So we have chosen b = 0.0001 and d = –0.001 for our calculation. As the value of $\lambda$ should be close to 0.5 in quite a broad range of low-x [13], we have taken $\lambda = 0.5$ in our work.

In figures 1(a) to 1(f), we compare our result of t-evolution of gluon distribution function from equation (15) with GRV '98 gluon distribution parameterization at x = $10^{-5}$ and $10^{-4}$ respectively. At both the x values we compare our results for $K(x) = k$, $ax^b$ and $ce^{dx}$. At x = $10^{-5}$, the best fit results are for k = 0.07, a = 0.07 and c = 0.07 and at x=$10^{-4}$, the best fit results are for k = 0.085, a = 0.085 and c = 0.085. The figures show good agreement of our result with GRV '98 parameterization at low-x.

In figures 2(a) to 2(c), we compare our result of x-evolution of gluon distribution function from equation (17) with MRST 2001 gluon distribution parameterization at $Q^2 = 20$ GeV$^2$ for $K(x) = k$, $ax^b$ and $ce^{dx}$ respectively. The best fit results are for k = 0.1, a = 0.1 and c = 0.1. As Regge theory strictly applicable only for low-x and high-$Q^2$ [13, 40], the best fits of our results with MRST 2001 are not so good. Figure 2(d) shows the comparison of our result of x-evolution of gluon distribution function given by equation (20) with MRST 2001 parameterization. Comparison of figure 2(d) with figures 2(a)–2(c) gives the contribution of quark to the gluon distribution function at x = 0.01 is ≈ 35%.

In figures 3(a) to 3(c), we compare our result of x-evolution of gluon distribution function from equation (17) with MRST 2004 gluon distribution parameterization at $Q^2 = 100$ GeV$^2$ for K (x) = k, $ax^b$ and $ce^{dx}$ respectively. The best fit results are for k= 0.16, a = 0.165 and c = 0.162. Figure 3(d) shows the comparison of our result of x-evolution of gluon distribution function given by equation (20) with MRST 2004 parameterization. Comparison of figure 3(d) with figures 3(a)–3(c) gives the contribution of quark to the gluon distribution function at x = 0.001 is ≈ 35% and at x = 0.01 is ≈ 30%. So we have seen from the above comparisons with MRST parameterizations that the contribution of quark decreases with increasing $Q^2$, and in our x-$Q^2$ region of discussion quark contributes appreciably to gluon distribution function. So we cannot ignore the contribution of quark in that region.

In figures 4 to 6 we compare our results of x-evolution of gluon distribution function from equations (17) and (20) with GRV '98 parameterization at $Q^2$ = 20, 40 and 80 GeV$^2$ respectively. We compare our results in the range $10^{-5} \leq x \leq 10^{-1}$ and we get a very good fit of our result to the GRV '98 parameterization. In some recent papers [41] Choudhury and Saharia, presented a form of gluon distribution function at low-x obtained from a unique solution with one single initial condition through the application of the method of characteristics [42]. They have overcome the limitations of non-uniqueness of some of the earlier approaches [15]. So, it is theoretically and phenomenologically favoured over the earlier approximations. We have presented these results with GRV '98 parameterizations and our results, and found that with decreasing x we get a better fit of our result to GRV '98 parameterization in comparison with those results.

Figures 4(a) to 4(c) are the best fit of our results to GRV '98 parameterization at $Q^2 = 20$ GeV$^2$ corresponding to k = 0.4, a = 0.4, c = 0.4. Figure 4(d) shows the comparison of the result of equation (20) with GRV '98 parameterization at $Q^2 = 20$ GeV$^2$, and we get the contribution of quark to the gluon distribution function at x = $10^{-5}$ is ≈ 7.5%. Figures 5(a) to 5(c) are the best fit of our results to GRV '98 parameterization at $Q^2 = 40$ GeV$^2$ corresponding to k = 0.7, a = 0.7, c = 0.7. Figure 5(d) shows the comparison of the result of equation (20) with GRV '98 parameterization at $Q^2 = 40$ GeV$^2$, and we get the contribution of quark to the gluon distribution function at x = $10^{-5}$ is ≈ 4.7%. Figures 6(a) to 6(c) are the best fit of our results to GRV '98 parameterization at $Q^2 = 80$ GeV$^2$ corresponding to k = 1, a = 1, c = 1. Figure 6(d) shows the comparison of the result of equation (20) with GRV '98 parameterization at $Q^2 = 80$ GeV$^2$, and we get the contribution of quark to the gluon distribution function at x = $10^{-5}$ is ≈3.5%. From the figures 4 to 6, we have seen that with the increase of $Q^2$, the quark contribution to the gluon distribution function decreases and the contribution at x = $10^{-5}$ is very small in the range $20 \leq Q^2 \leq 80$ GeV$^2$, which is clear from GRV '98 parameterization of gluon distribution function. So the contribution of quark to the gluon distribution function can be ignored for sufficiently low-x and high -$Q^2$ range.

Figures 7(a) to 7(f) show the sensitivity of the parameters λ, k, a, b, c and d respectively. Taking the best fit figure to the x-evolution of gluon distribution function of MRST 2004 parameterization at $Q^2 = 100$ GeV$^2$, we have given the ranges of the parameters as 0.4≤λ≤0.6, 0.12≤k≤0.2, 0.12≤a≤0.26, 0.0001≤b≤0.02, 0.12≤c≤0.26 and −2.5≤d≤−0.001.

**4 Conclusions**

We have considered the Regge behaviour of singlet structure function and gluon distribution function to solve DGLAP evolution equations. Here we find the t and x-evolutions of gluon distribution function in LO. We see that our results are in good agreement with MRST and GRV '98 global parameterizations especially at low-x and high-$Q^2$ region. We can conclude that Regge behaviour of quark and gluon distribution functions is compatible with PQCD at that region assuming the Regge intercept to be the same for both quark and gluon. We have also seen from GRV '98 parameterization that the quark contribution to the gluon distribution function can be neglected at $x \leq 10^{-5}$ and $Q^2 \geq 80$ GeV$^2$, as in this range the contribution becomes less than 4%. Here we have overcome the limitations that arise from Taylor series expansion method. Considering Regge behaviour of distribution functions DGLAP equations become quite simple to solve and so this method is a viable simple alternative to other methods. But here also the problem of ad hoc assumption of the function K(x), the relation between singlet structure function and gluon distribution function could not be overcome. Work is going on to obtain the simultaneous solution of coupled DGLAP evolution equations for singlet structure function and gluon distribution function. Moreover here we solve only leading order evolution equations. We expect that next-to-leading order equations are more correct and their solutions will give better fit to global data and parameterizations.

*Acknowledgements.* We are grateful to G. A. Ahmed for his help in numerical part of this work. One of us (JKS) is grateful to the University Grants Commission, New Delhi for the financial assistance to this work in the form of a major research project.

**Appendix**

```
//*t- evolution of gluon structure function for k=constant*//
#include<math.h>
#include<stdio.h>
#include<conio.h>
#define Nf      4
#define t       (log(Q²)-log(Z²))
#define t₀      (log(Q₀²)-log(Z²))
#define Af      0.16
#define C       0.694444
```

```c
#define Z        0.323
#define K        0.11
#define λ        0.5
#define x        0.01
#define Q_0^2    10
#define G(x_0,t) 5.67
#define div      10000
#define ul       0.9999
#define f(w)     (K*((pow(w,(λ+1))-1)/(1-w))+(1-w)*(w+pow(w,-1))*K*pow(w,λ)+0.222222*((1+pow((1-w),2))/w)*pow(w,λ))
#define p        (U/K)
#define B        (Af*(C+log(1-x)+p))
#define G(x,t)   (G(x_0,t)*(pow((t/t_0),B)))
main()
{
int  i;
double  h,s,sa,sb,U,Q^2;
clrscr();
h=(ul-x)/div;
s=f(ul)+f(x);
for(i=1;i<=div-1;i=i+1)
{
sa=(x+(i*h));
s=s+(2*(f(sa)));
}
for(i=1;i<=div-1;i=i+2)
{
sb=(x+(i*h));
s=s+(2*(f(sb)));
}
U=(s*h)/3.0;
printf("\n integral=%lf", U);
printf("\n value of p=%lf", p);
printf("\n value of Q^2:   ");
scanf("%lf",  &Q^2);
printf("\n value of G(x,t)=%lf", G(x,t));
getch();
return(0);
}

//*x-evolution of gluon structure function for k=constant*//
#include<math.h>
#include<stdio.h>
#include<conio.h>
#define N_f      4
#define A_f      0.16
#define C        0.694444
#define Z        0.323
#define K        0.16
#define λ        0.6
#define x_0      0.1
#define Q^2      100
#define G(x_0,t) 1.1765
#define div      1000
```

```c
#define ul       0.9999
#define f(w)     (K*((pow(w,($\lambda$+1))-1)/(1-w))+(1-w)*(w+pow(w,-1))*K*pow(w,lam)+0.222222*((1+pow((1-w),2))/w)*pow(w,$\lambda$))
#define p        (U/K)
#define B        ($A_f$*(C+log(1-x)+p))
#define $p_0$    ($U_0$/K)
#define $B_0$    ($A_f$*(C+log(1-$x_0$)+$p_0$))
#define G(x,t)   (G($x_0$,t)*(pow((log($Q^2$)-log($Z^2$)),(B-$B_0$)))*(pow(($x_0$/x),$\lambda$)))
main()
{
int  i;
double  h,s,sa,sb,U,$h_0$,$s_0$,$sa_0$,$sb_0$,$U_0$,x;
clrscr();
h0=(ul-$x_0$)/div;
s0=f(ul)+f($x_0$);
for(i=1;i<=div-1;i=i+1)
{
$sa_0$=($x_0$+(i*$h_0$));
$s_0$=$s_0$+(2*(f($sa_0$)));
}
for(i=1;i<=div-1;i=i+2)
{
$sb_0$=($lm_0$+(i*$h_0$));
$s_0$=$s_0$+(2*(f($sb_0$)));
}
$U_0$=($s_0$*$h_0$)/3.0;
printf("\n integral=%lf", $U_0$);
printf("\n value of p0=%lf", $p_0$);
printf("\n value of x:   ");
scanf("%lf",  &x);
h=(ul-x)/div;
s=f(ul)+f(x);
for(i=1;i<=div-1;i=i+1)
{
sa=(x+(i*h));
s=s+(2*(f(sa)));
}
for(i=1;i<=div-1;i=i+2)
{
sb=(x+(i*h));
s=s+(2*(f(sb)));
}
U=(s*h)/3.0;
printf("\n integral=%lf", U);
printf("\n value of p=%lf", p);
printf("\n value of G(x,t)=%lf", G(x,t));
getch();
return(0);
}
```

**Figure captions**

**Fig. 1.** t-evolution of gluon distribution function in LO for $\lambda = 0.5$ and $K(x) = k$, $ax^b$ and $ce^{dx}$ for the representative values of x. Data points at lowest-$Q^2$ values are taken as input to test the evolution equation (15). Here Fig. 1(a) to 1(f) are the best fit graphs of our results with GRV '98 parameterization corresponding to $K(x) = k$, $ax^b$ and $ce^{dx}$ respectively.

**Fig. 2.** x-evolution of gluon distribution function in LO for $\lambda = 0.5$ and $K(x) = k$, $ax^b$ and $ce^{dx}$ for $Q^2 = 20$ GeV$^2$. Data points for x values just below 0.1 are taken as input to test the evolution equation (17) and (20). Here Fig. 2(a) to 2(d) are the best fit graphs of our result with MRST 2001 parameterization.

**Fig. 3.** Same as Fig. 2 for $Q^2 = 100$ GeV$^2$. Here Fig. 3(a) to 3(d) are the best fit graphs of our result with MRST 2004 parameterization.

**Fig. 4.** Same as Fig. 2 for $Q^2 = 20$ GeV$^2$. Here Fig. 4(a) to 4(d) are the best fit graphs of our result with GRV '98 parameterization.

**Fig. 5.** Same as Fig. 4 for $Q^2 = 40$ GeV$^2$.

**Fig. 6.** Same as Fig. 4 for $Q^2 = 80$ GeV$^2$.

**Fig. 7.** Fig. 7(a) to 7(f) show the sensitivity of the parameters $\lambda$, k, a, b, c and d respectively at $Q^2 = 100$ GeV$^2$.

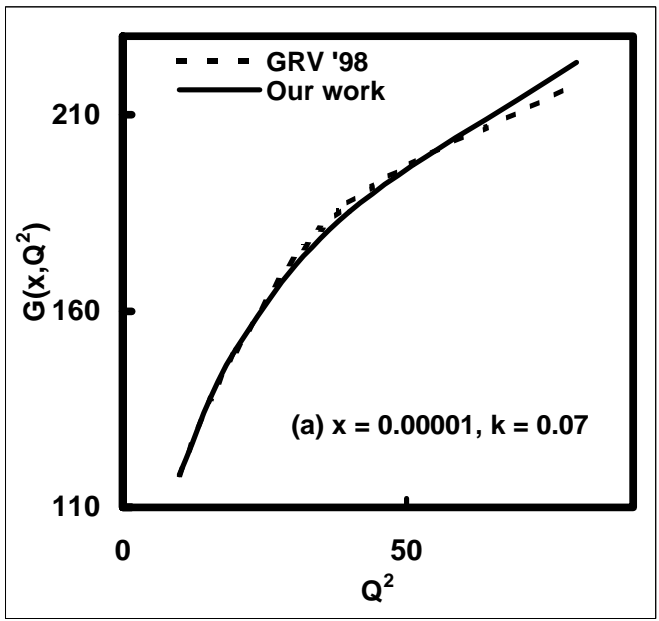
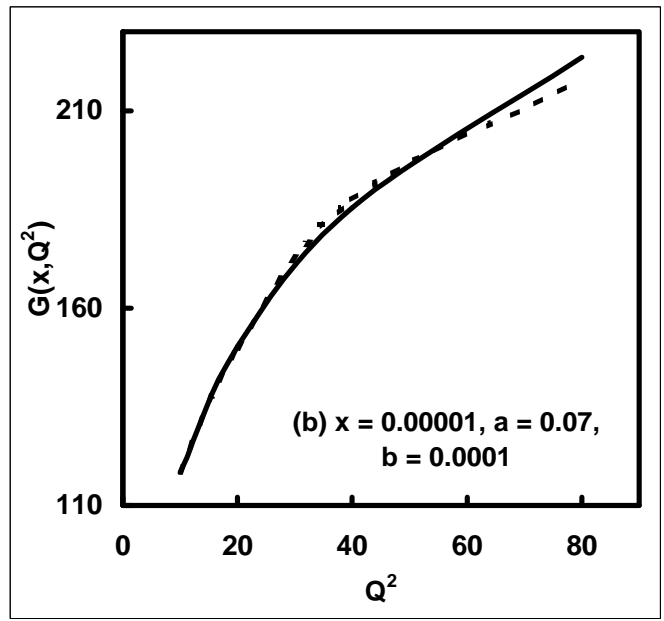
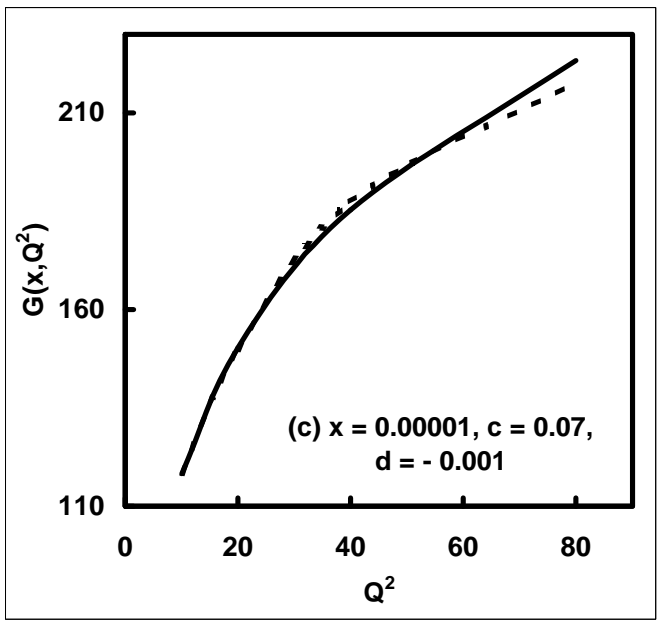
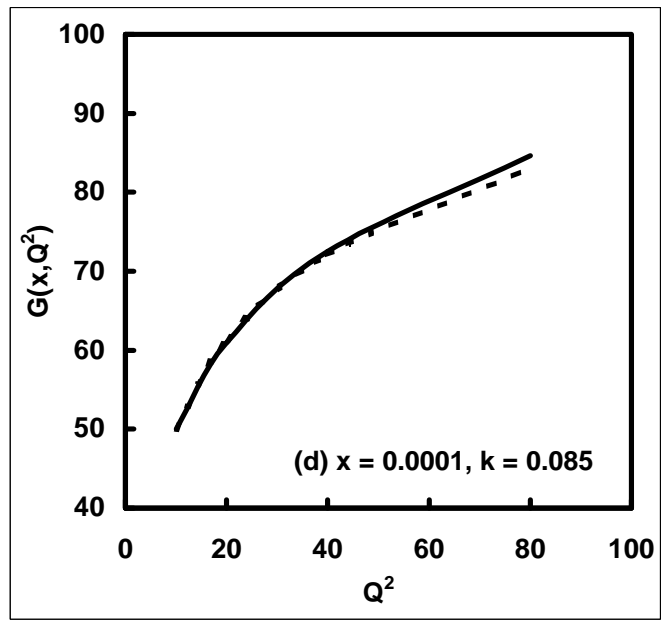
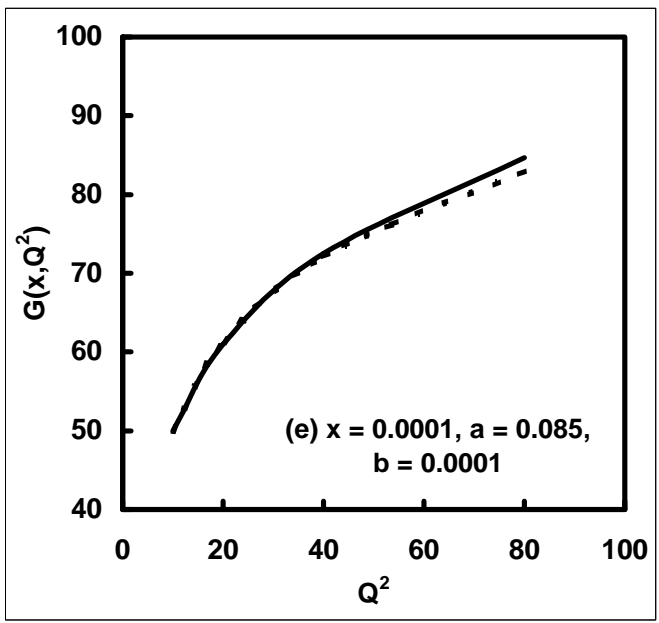
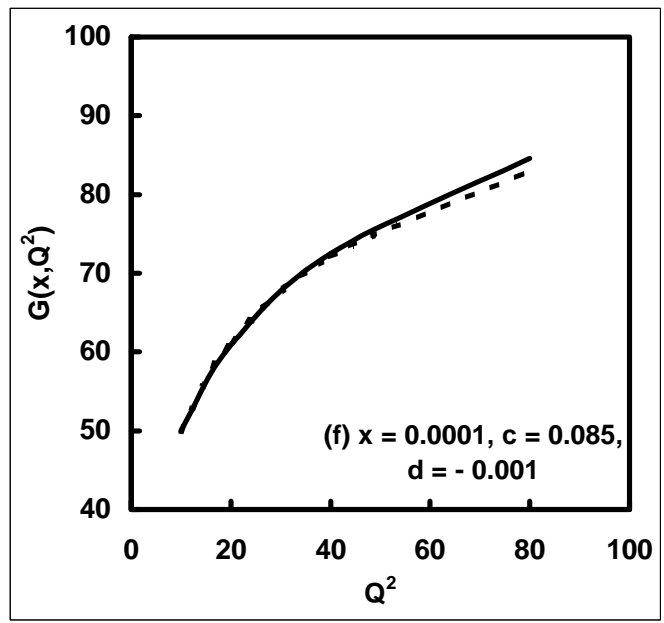

**Fig.1**

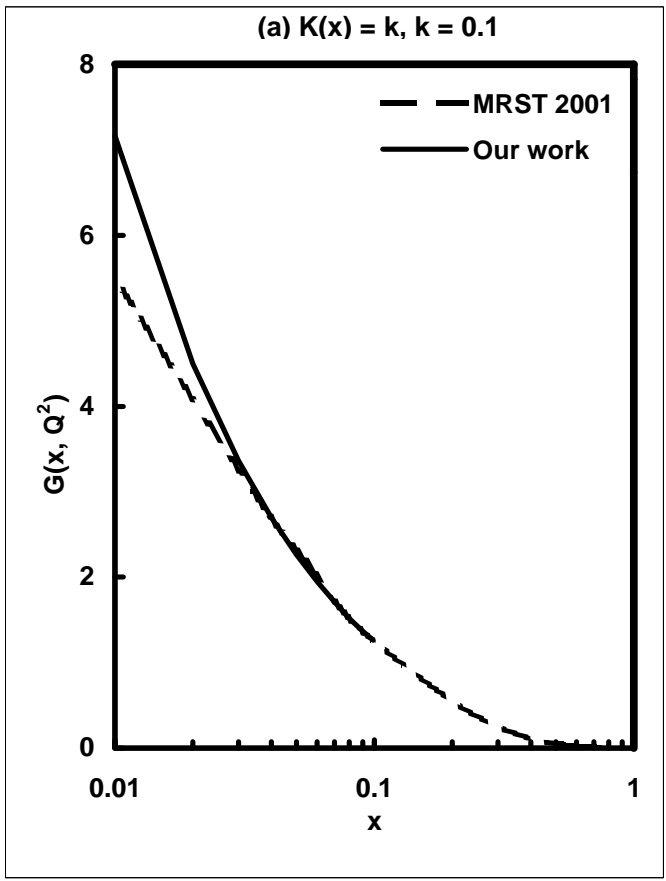
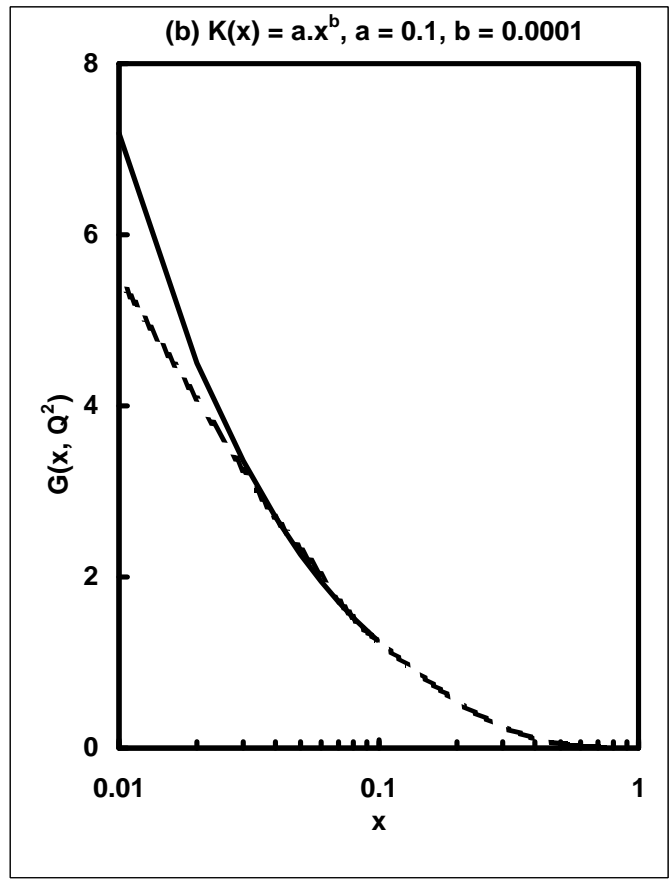
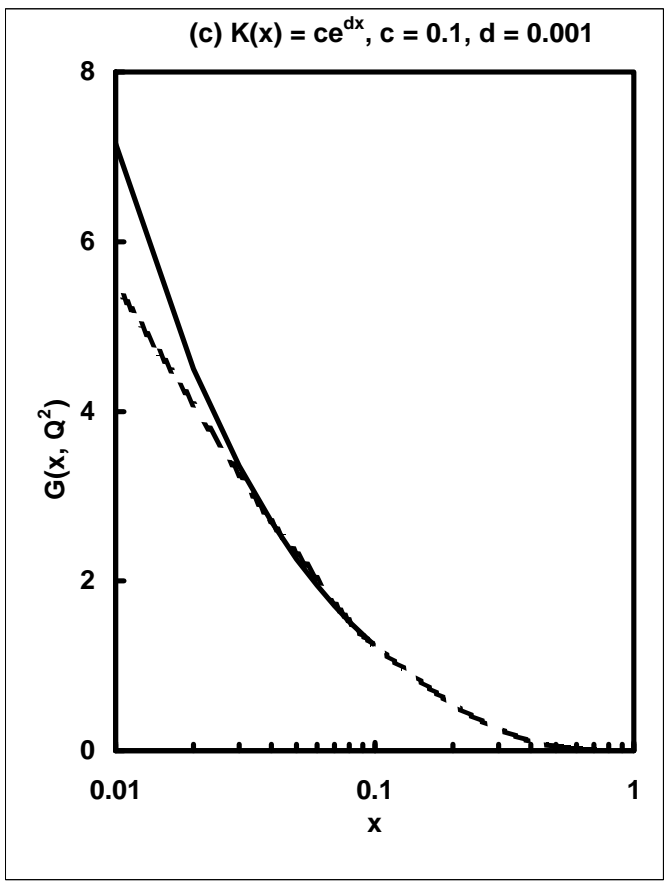
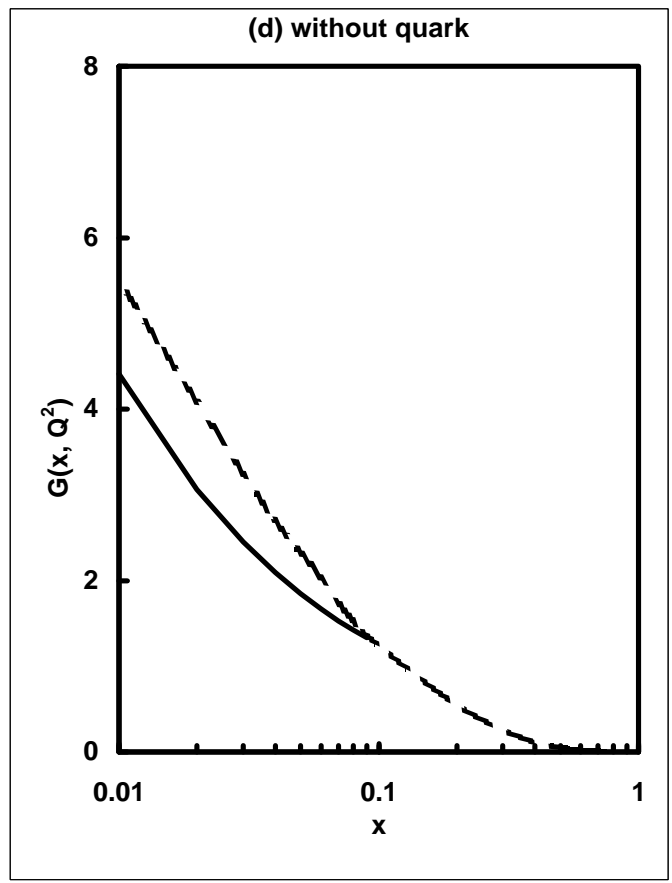

**Fig.2**

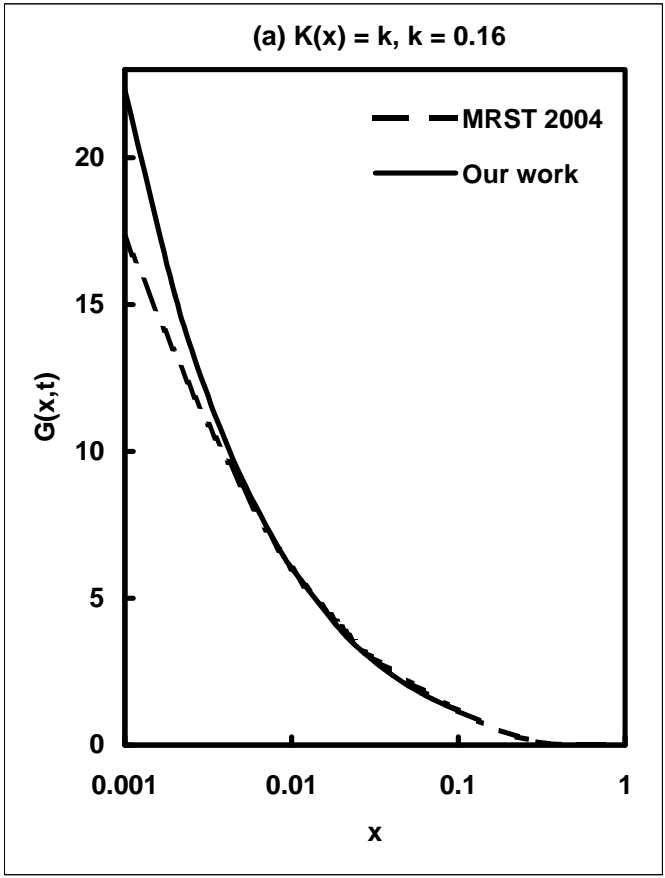
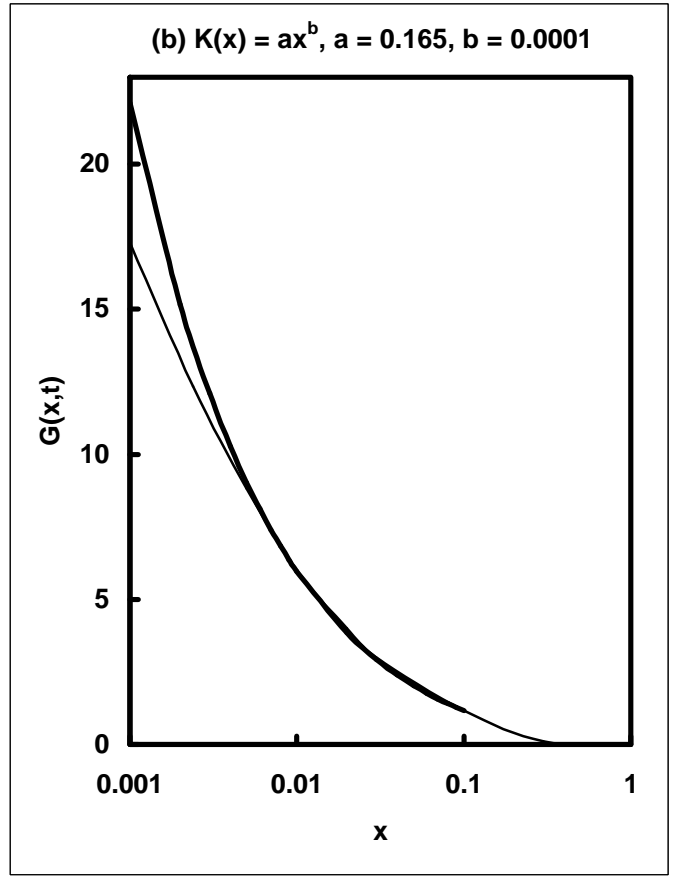
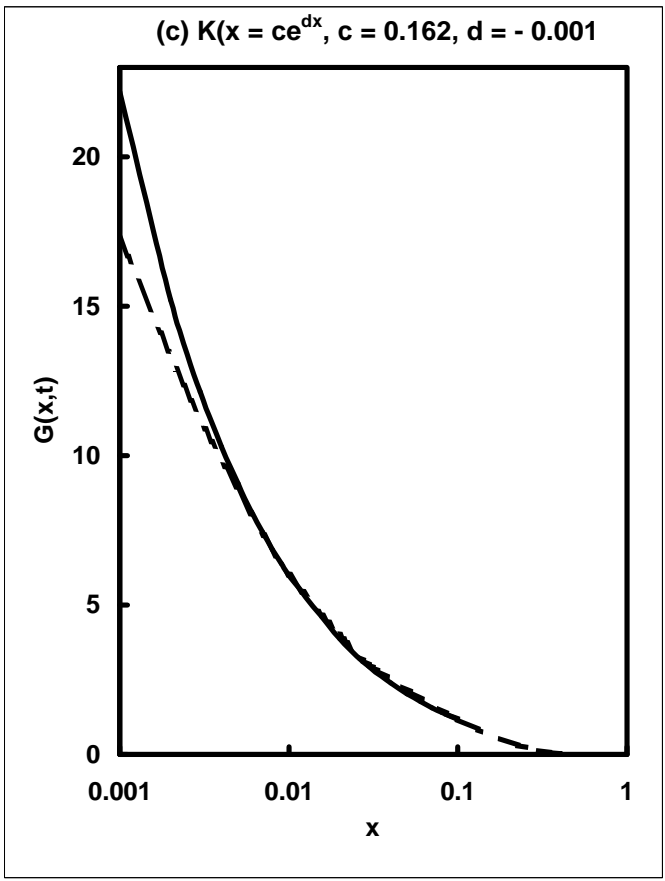
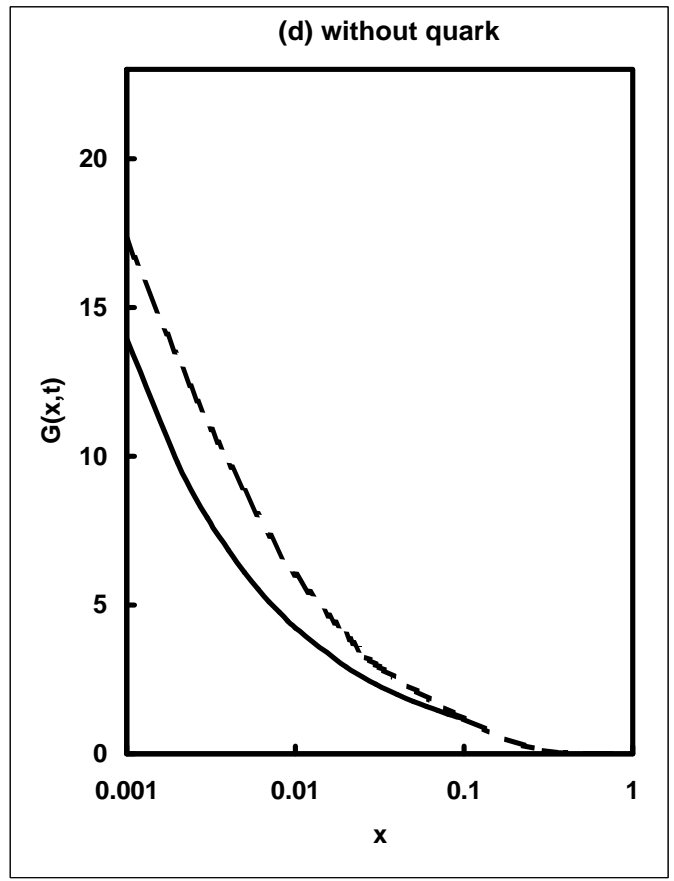

**Fig.3**

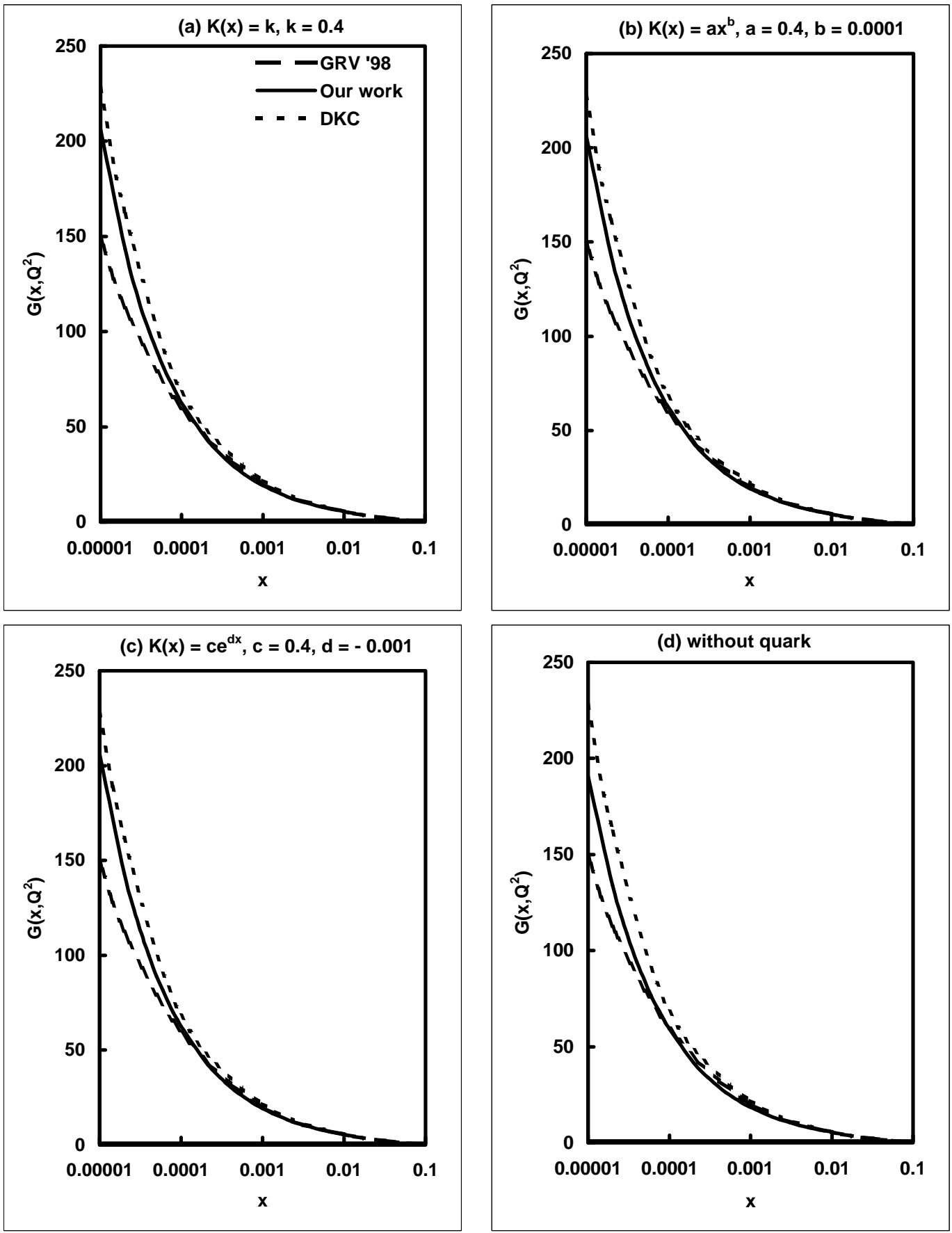

**Fig.4**

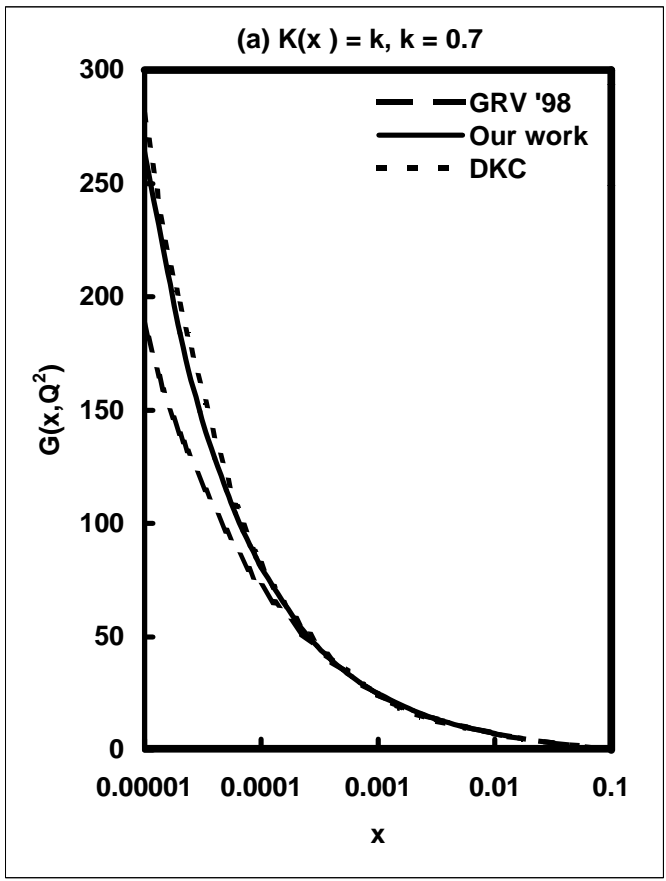
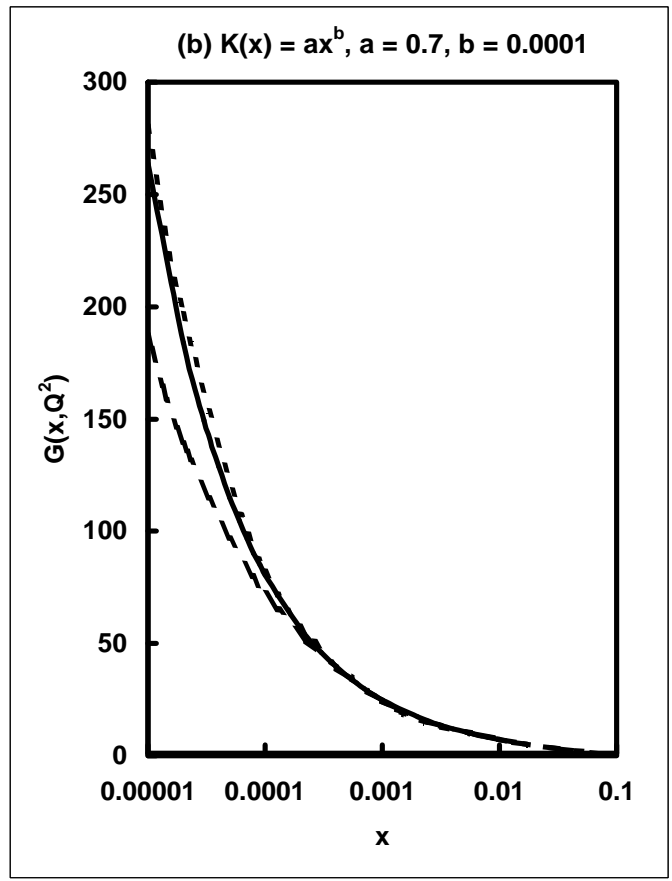
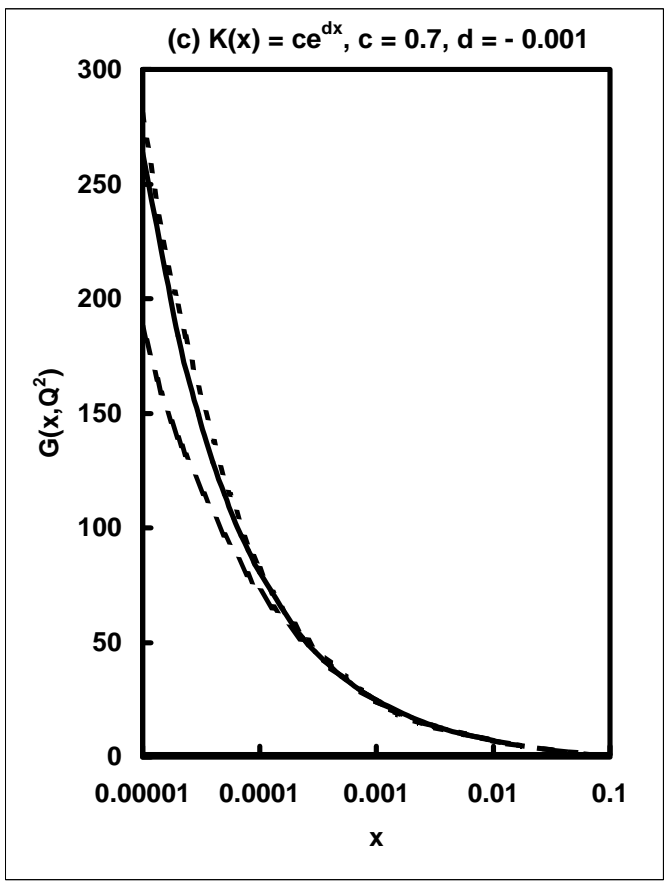
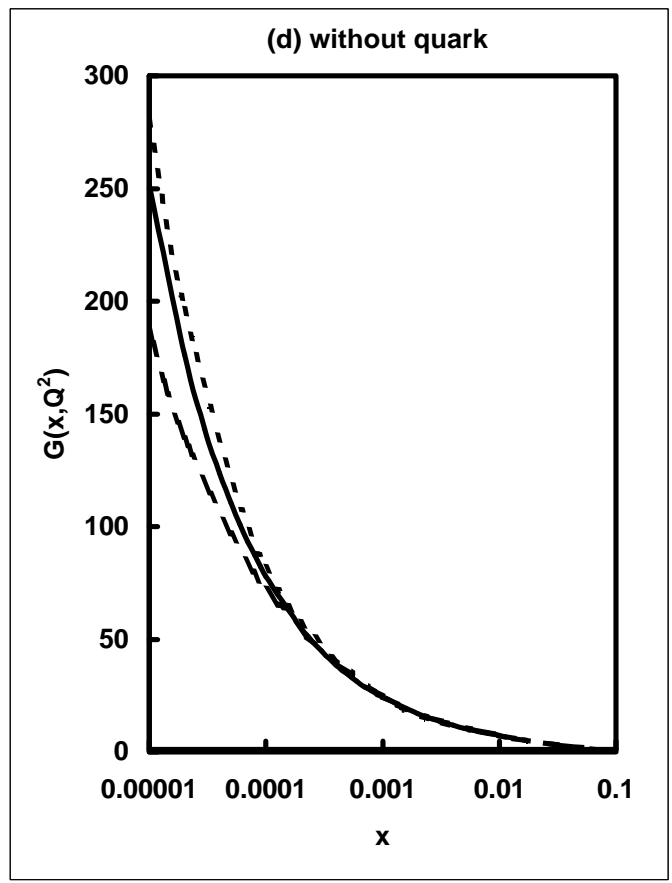

**Fig.5**

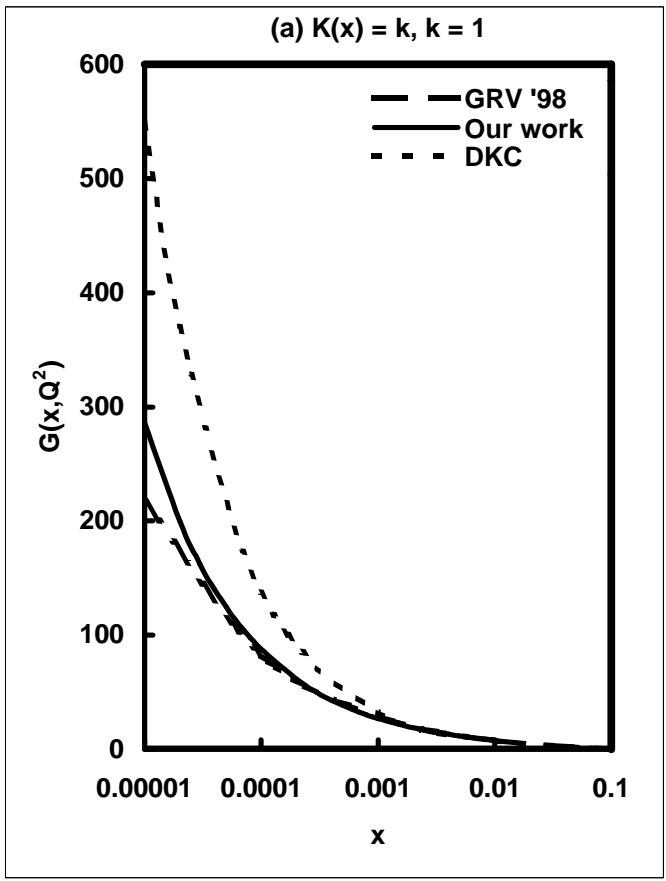
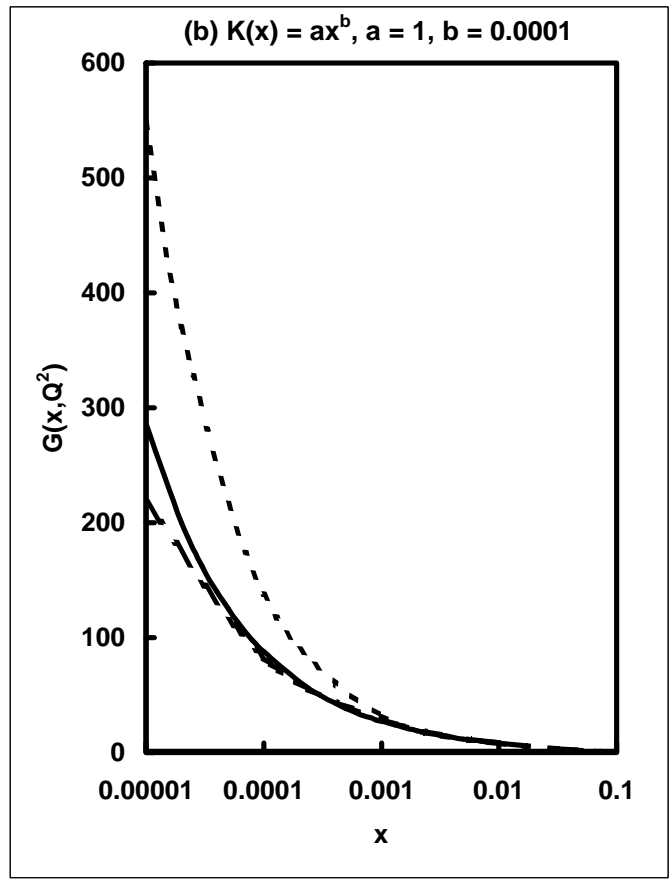
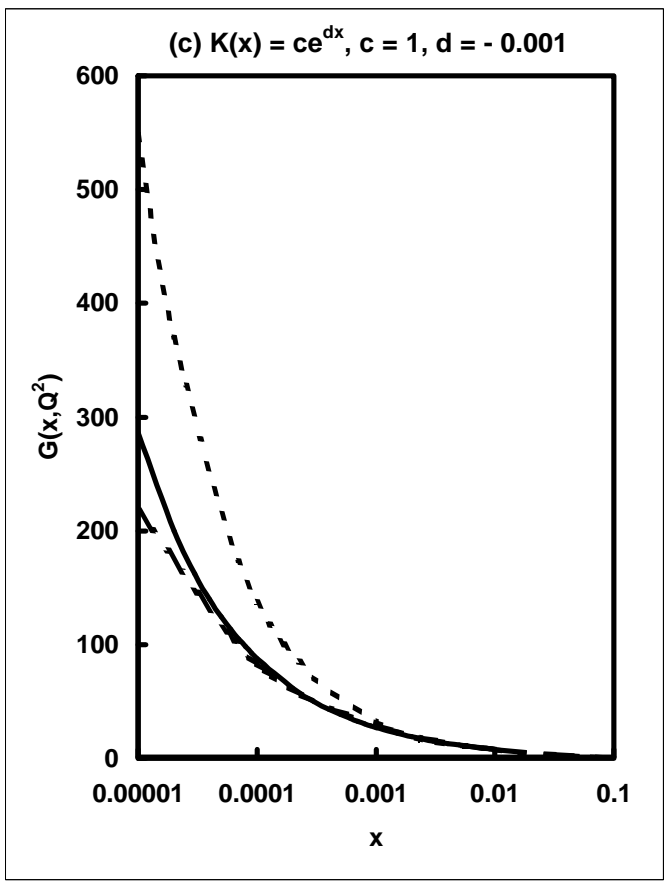
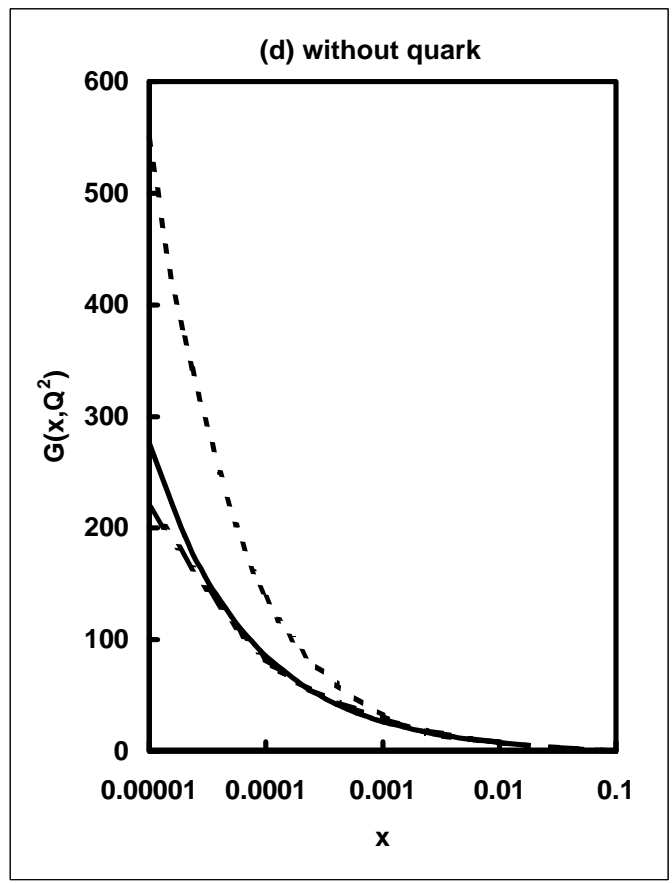

**Fig.6**

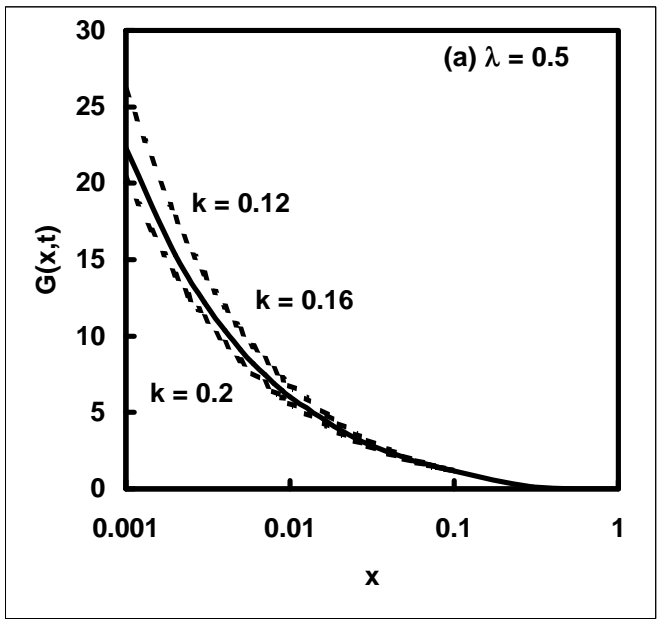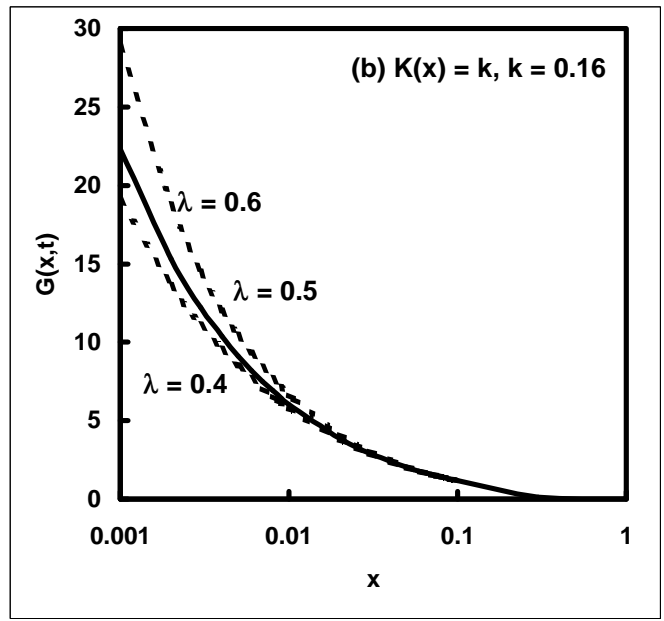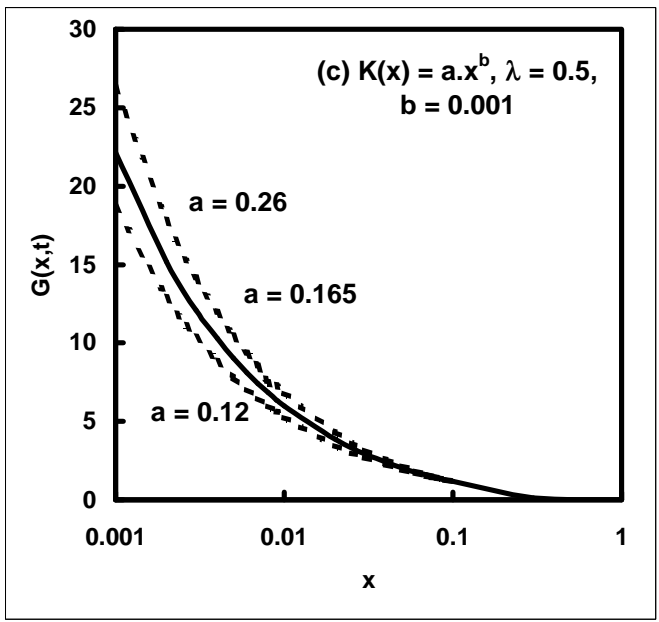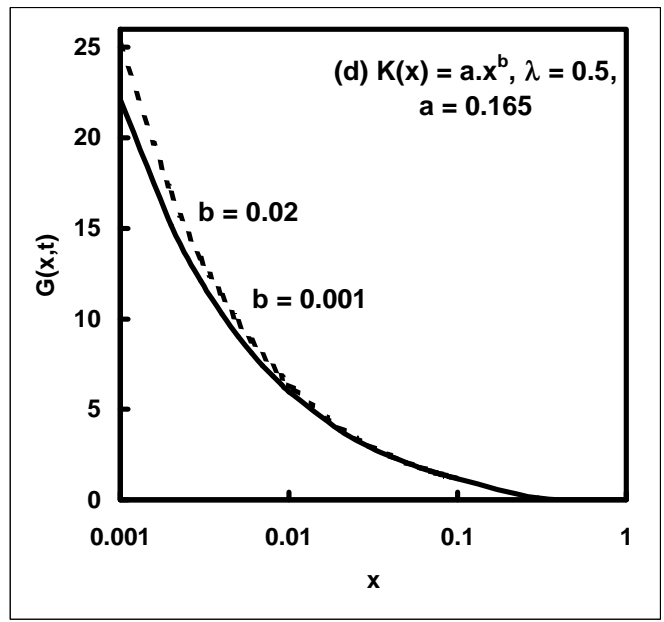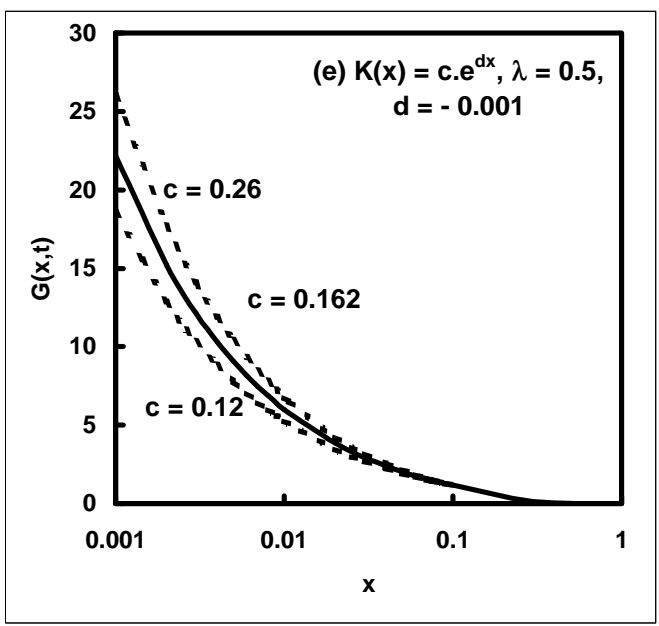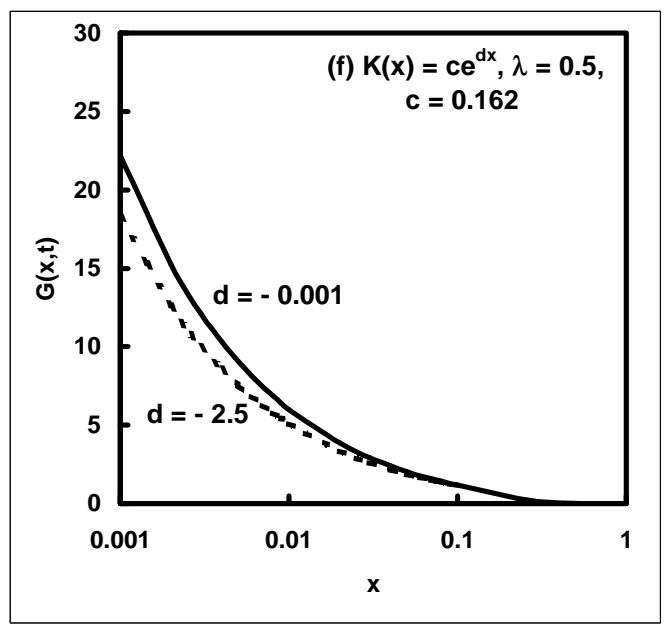

**Fig.7**